\documentclass[showpacs,aps,prb,twocolumn,superscriptaddress]{revtex4-1}
\usepackage{graphicx} 
\usepackage{dcolumn}
\usepackage{bm}
\usepackage{amssymb,amsmath}
\usepackage{epstopdf}
\usepackage{xcolor} 
\usepackage[normalem]{ulem} 
\begin{document}
\title{Polarization Selective Modulation of the Supercavity Resonance from Friedrich-Wintgen Bound States in the Continuum}
\normalsize
\author{C. Kyaw}
\affiliation{Department of Physics and Astronomy, Howard University, Washington, DC 20059, USA}
\author{R.~Yahiaoui}
\affiliation{Department of Physics and Astronomy, Howard University, Washington, DC 20059, USA}
\author{J. A.~Burrow}
\affiliation{Electro-Optics Department, University of Dayton, Dayton, OH, 45469, USA}
\author{V.~Tran}
\affiliation{Department of Physics and Astronomy, Howard University, Washington, DC 20059, USA}
\author{K.~Keelen}
\affiliation{Department of Physics $\&$ Dual-Degree Engineering, Morehouse College, 
Atlanta, GA, 30314}
\author{W.~Sims}
\affiliation{Department of Physics $\&$ Dual-Degree Engineering, Morehouse College, 
Atlanta, GA, 30314}
\author{E. C.~Red}
\affiliation{Department of Physics $\&$ Dual-Degree Engineering, Morehouse College, 
Atlanta, GA, 30314}
\author{W. S.~Rockward}
\affiliation{Department of Physics , Morgan State University, Baltimore, MD, USA}
\author{M. A.~Thomas}
\affiliation{Institute for Electronics and Nanotechnology, Georgia Institute of Technology, Atlanta, GA 30332, USA}
\author{A.~Saragan}
\affiliation{Electro-Optics Department, University of Dayton, Dayton, OH, 45469, USA}
\author{I.~Agha}
\affiliation{Electro-Optics Department, University of Dayton, Dayton, OH, 45469, USA}

\author{T. A.~Searles}
\email[]{thomas.searles@howard.edu}
\thanks{corresponding author.}
\affiliation{Department of Physics and Astronomy, Howard University, Washington, DC 20059, USA}
\date{\today}

\begin{abstract}
An article usually includes an abstract, a concise summary of the work
covered at length in the main body of the article. 
\begin{description}
\item[Usage]
Secondary publications and information retrieval purposes.
\item[Structure]
You may use the \texttt{description} environment to structure your abstract;
use the optional argument of the \verb+\item+ command to give the category of each item. 
\end{description}
\end{abstract}
\begin{abstract}
Bound states in the continuum (BICs) are widely studied for their ability to confine light, produce sharp resonances for sensing applications and serve as avenues for lasing action with topological characteristics. Recent experiments have demonstrated the existence of exotic modes which occur in off-$\Gamma$\ points not accessible by symmetry-protected BICs, Freidrich-Wintgen (FW) BICs. In previous works,  FW BICs were formed by either varying the incident angle or through geometric manipulation. On the contrary, in this work we demonstrate the formation of FW BICs induced by different linear polarization states of incident terahertz waves.  Furthermore, as predicted by  theory, the position of the FW BIC region is verified to change as a function of asymmetry. This allows for a unique parameter, incident polarization, to induce the FW BIC and modulate the supercavity resonance with the degree of asymmetry. 
The polarization-selective quasi-BICs observed near the FW BIC have potential applications in lasing, spectral filtering and high-performance sensing devices.
\end{abstract}

\maketitle

\section{\label{sec:level1}Introduction\vspace{-1em}}
The concept of bound states in the continuum, or BIC, was first proposed by von Neumann and Wigner for an electron in an artificial complex potential \cite{von1929}. Thereafter, different types of BICs have been reported in quantum systems\cite{Schult89,Exner1996}, acoustic and water waves \cite{Parker66,ursell51} and photonic systems, \cite{Watts:02,Marinica08,Molina12,Monticone14,Hsu2013,Jin2019,Shuyuan2019} where modes inside the radiation continuum above the light line are perfectly confined instead of radiating away. Due to the infinite quality (Q-) factor and zero linewidth of the BIC modes, only quasi-BICs with partial confinement and finite linewidth are observed experimentally. The partial confinement of waves has applications in lasing \cite{Rybin2017,Ha2018,Kodigala2017}, nonlinear phenomena \cite{Liu2012,Koshelev2018,Vabishchevich2018} and high-performance sensing devices. \cite{Yahiaoui2015,Evlyukhin2010,Liu:17}

In general, bound states in the continuum originate from two physical mechanisms. The more common type, symmetry-protected BIC results from a symmetry mismatch between radiative modes and the mode profiles inside the Brillouin zone near the $\Gamma$\ point. The second classification, Friedrich-Wintgen (FW) BICs (or accidental BICs), originates from  destructive interference between two radiation modes modulated by a specific parameter \cite{Freidrich1985}. Unlike symmetry-protected BICs,  FW BICs are observed at off-$\Gamma$\ points in regions with an accidental symmetry. These regions of the band dispersion result in extremely high Q-factors around the FW BIC, called near-BIC \cite{Azzam2018} or supercavity resonances,\cite{Rybin2017, Han2018} and are attributed to the coupling strength of the two radiative modes that interfere destructively.  The position and Q-factor of the supercavity resonances can be extremely sensitive to structural parameters of the device \cite{Kikkawa2019}. Furthermore, in accordance with FW BIC theory, changes in the coupling strength of the  resonances result in a shift of the FW BIC positions in the band dispersion\cite{Feshbach1958}. Efficient control of the  position and Q-factor of the supercavity resonance is desirable for practical design of high Q-factor devices.

Limited experimental demonstrations of FW BICs have been shown to be dependent on  angles of incident radiation in the infrared regime \cite{Hsu2013,Yoon2015, Azzam2018}. At terahertz (THz) frequencies, symmetry protected BICs \cite{SinghBIC,Abujetas2019,Fan2019} with few realizations of FW BICs \cite{Han2018,Zhang2018} have been reported. Recent investigations by Han \textit{et al}. \cite{Han2018} reported resonance-trapped BICs with similar mechanism as FW BICs for THz metasurfaces by tailoring the geometric lengths of the silicon resonators. They also reported frequency and Q-factor modulation of the supercavity resonance through optical pumping of the silicon resonators.

In this Letter, we report a novel method, altering the linear polarization state of the incident THz wave, to control the coupling strength of the interfering resonances that produce FW BICs. Through this method, we modulate the position and bandwidth of FW BICs in asymmetric terahertz metasurfaces. Previous works induce a FW BIC through non-zero incident angles or tailoring the geometry of the device. The results of this work provide an active methodology to manipulate the extremely high Q-factor resonances of FW BICs via manipulation of the incoming radiation which could be advantageous for photonic, lasing and sensing applications. The paper is organized as follows. First, we confirm that a quasi-BIC is formed by breaking the symmetry of the SRR unit cell.  Next, we investigate the influence of polarization on the transmission spectra of the terahertz metasurface at highest asymmetry, where a FW BIC region is formed between $26^{\circ}$ $\leq \theta \leq$ $36^{\circ}$. Finally, we show that the spectral position and Q factor of the supercavity resonances around this FW BIC region can be precisely tuned as a function of structural asymmetry.

\section{\label{sec:level1}Materials and Methods}

\begin{figure}[ht!]
\centering\includegraphics[width=8cm]{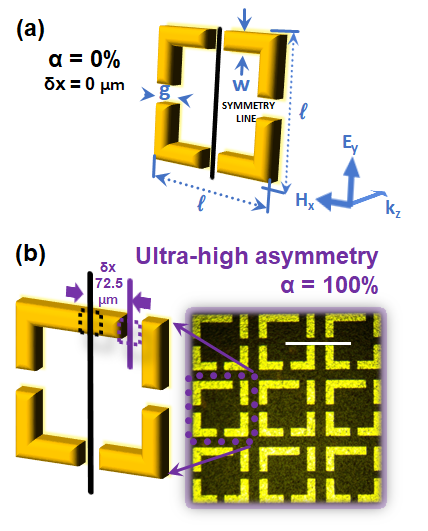}
\caption{Schematic of unit cell geometry of the asymmetric THz metasurface array. (a) Geometrical description of the symmetric SRR unit cell with $l$ = 250 $\mu$m, $w$ = 35 $\mu$m, and $g$ = 35 $\mu$m, (b) unit cell geometry for sample with $\delta$ = 72.5 $\mu$m or $\alpha$ = 100\% and its respective optical image. Scale bar: 300 $\mu$m.}
\label{fig:geometry}
\end{figure}

The unit cell geometry of a symmetric SRR with four capacitive gaps at $\delta$ = 0 $\mu$m is shown in Fig.~1(a).  The dimensions are length $l$ of 250 $\mu$m, bracket width $w$ of 35 $\mu$m and gap width $g$ of 35 $\mu$m arranged in an array with 300 $\mu$m period. 
An asymmetry parameter, $\alpha$ has been defined as 100\% $\times$ $\delta_{i}$/$\delta_{max}$ where $\delta_{i}$ is the distance between the gap center and the symmetry line and $\delta_{max}$ is 72.5 $\mu$m as shown in Fig. 1(b).
Experimentally, the SRR arms consist of 100 nm thick silver layer deposited by RF sputtering on polyimide substrate of 50.8 um thickness using conventional lithography.  An optical microscope image of the metasurface is also shown in Fig.~1(b). 

A high-resolution continuous wave THz spectrometer  (Teraview CW Spectra 400) that produces linearly polarized collimated terahertz beams was utilized to measure the transmission spectra of the aforementioned fabricated metasurfaces. In this system, the optical beat frequency of two  distributed feedback near-IR diode lasers are  tuned to produce coherent THz waves between 0.05 to 1.5 THz. The spectral resolution of around 100 MHz is enough to resolve relatively narrow spectral features, in contrast to the limited resolution of time-domain terahertz systems. 

Samples were placed equidistant between the  emitter and detector in ambient air conditions. After measurement, the transmission spectrum was calculated as $T(f) = P_{M}(f)/P_{sub}(f)$, where $P_{M}(f)$ and $P_{sub}(f)$ are the filtered power spectra of the metasurface and substrate, respectively. 
For the polarization dependence investigation, the metamaterial sample is rotated by a value $\theta$ to mimic the change of a polarization state of the incident beam \cite{Burrow17}.

For numerical simulations, we utilized  a finite element method with periodic boundary conditions to simulate a 2D infinite array of unit cells.  The length scale of the mesh was set to be less than or equal to $\lambda_{0}$ /10 throughout the simulation domain, where $\lambda_{0}$ is the central wavelength of the incident radiation. The input and output ports are located at  3$\lambda_{0}$ from the metasurface with open boundary conditions. The material parameters used for the structures are $\sigma$ = 4.1 x $10^7$ S/m for the metallic Ag layer and $\epsilon$ = 3.8 for the dielectric Kapton layer.
For the polarization dependent study, each unit cell was excited with a THz plane wave of the correct linear polarization state as indicated by the polarization angle from $0^{\circ}$ to $90^{\circ}$.

\section{\label{sec:level1}Results and Discussion}

Symmetry-protected BICs can be verified by breaking the structural symmetry of the device. Here, we experimentally show the quasi-BICs originating from the translation of the top gap of the unit cell, confirming the results of Cong \textit{et al}.\cite{SinghBIC}. In general, any small perturbation of the BIC gives rise to quasi-BICs that can be experimentally verified such that symmetry-protected bound states can radiate through coupling with the incoming radiation. 

\begin{figure}[h!]
\centering\includegraphics[width=\linewidth]{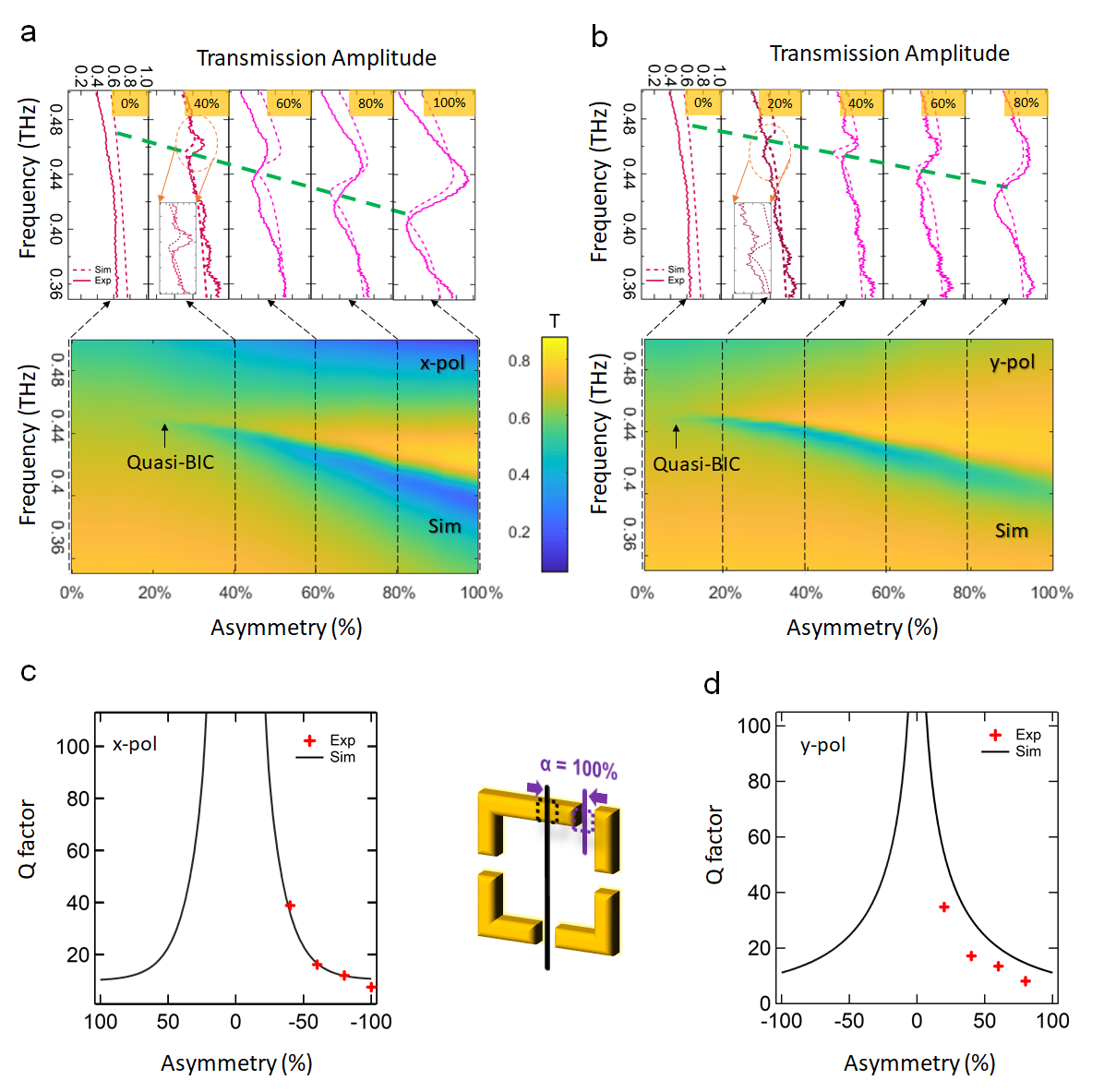}
\caption{Experimental and numerical investigation of the symmetry-protected BICs from quasi-BICs induced from top gap translation as illustrated in figure insert. Transmission spectra for (a) horizontal polarization, (b) vertical polarization with respective pictorial illustrations with dependence on asymmetry parameter $\alpha$ to exhibit the BIC nature from disappearance of the Fano at symmetric region. The Q factor for (c) x-pol and (d) y-pol BIC resonances as a function of asymmetry parameter $\alpha$ also confirm the diverging trend near the symmetric region.}
\end{figure}

\begin{figure}[h!]
\centering\includegraphics[width=\linewidth]{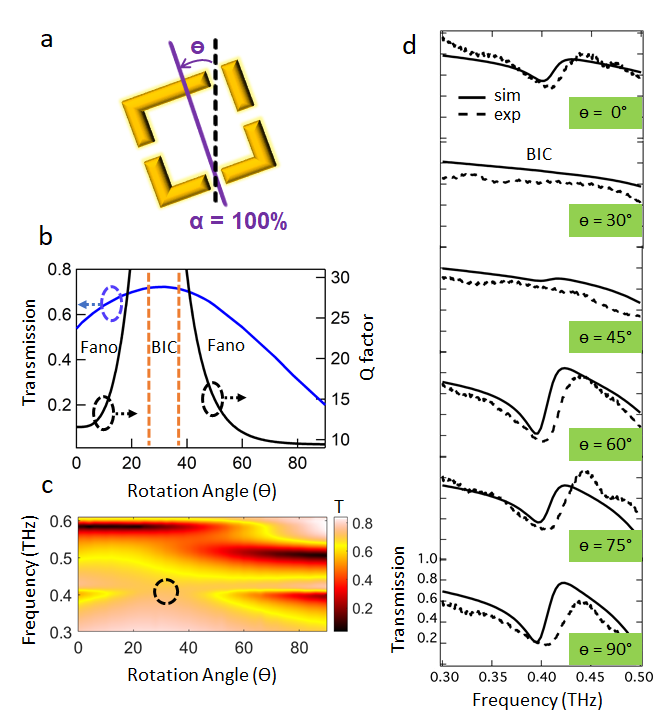}
\caption{Experimental and numerical investigation on polarization dependence of Fano resonance near 0.4 THz for the ultra-high asymmetry sample. (a) Sample $\alpha$ = 100\% is rotated counter-clockwise from  $\theta$ = $0^{\circ}$ to $\theta$ = $90^{\circ}$. (b) The transmission amplitude and Q-factor of the Fano resonance near 0.4 THz calculated from the simulated transmission spectra are plotted as a function of rotation angle. The plateau region in transmission indicates the FW BIC region induced by rotation. (c) Pictorial illustration of the transmission amplitude for different rotational angles between 0.3 and 0.6 THz with the dotted circle indicating the FW BIC region with vanishing linewidth. (d) Experimental and simulated  transmission spectra for different rotational angles $\theta$ corresponding to different linear polarization states confirm the complete disappearance of the Fano at $\theta$ = $30^{\circ}$ due to FW BIC.}
\end{figure}

As shown in the insert of Fig.~2, the top gap is translated laterally toward the right to break the $C_{4}$ symmetry and produce quasi-BICs in the form of a Fano resonance lineshape~\cite{Singh:11,Cong:15,Manjappa15,Srivastava15, Fedotov07, Singh:10, Singh11a}. The asymmetry parameter $\alpha$, definied previously, describes a general parameter common to all symmetry-protected BICs. The simulated and experimental transmission spectra are plotted when the incident radiation is horizontally polarized (x-pol) in Fig.~2(a) versus vertically polarized in Fig.~2(b). The transition from a symmetry-protected BIC to a quasi-BIC, due to asymmetry, is observed for both experiment and simulation in x- and y-polarizations. The degree of asymmetry before quasi-BICs can be detected is larger for x-polarization where the transmission dip only becomes visible in simulations around ~11\% as compared to ~3\% asymmetry for y-polarization. The extremely sharp quasi-BICs produced by very small perturbations are not resolved experimentally due to limitation imposed by the detection equipment. The quasi-BIC closest to symmetry-protected BIC modes are observed experimentally at $\alpha$ = 40\% for x-polarization and at $\alpha$ = 20\% for y-polarization. The inserts inside transmission graphs plotted in Figs.~2(a) and (b) show the quasi-BICs in more detail. The spectral feature of quasi-BIC also broadens with increasing asymmetry showing a strong dependence on asymmetric gap distance.

Further evidence of symmetry-protected BIC is represented by the disappearance of a resonance at the symmetry restoring region, where the linewidth becomes zero or quality factor tends to infinity~\cite{Hsu2013,Koshelev}. The Q-factors of the resonances are extracted as $Q = f_{i}$ /$\Delta$$f_{i}$, where $f_{i}$ is the center frequency and $\Delta$$f_{i}$ is the FWHM.  In Figs.~2(c) and (d), the calculated Q-factors of quasi-BICs show a clear diverging trend when the structure approaches symmetry. However, the Q-factor diverges closer to the symmetric region for y-polarization. This shows the incident field is more sensitive to the asymmetry in the structure when the electric field is orientated perpendicular to the translated gap. The experimentally measured Q-factor values are resolved for higher degree of asymmetry with 38.7 at $\alpha$ = 40\% for x-polarization and 34.8 at $\alpha$ = 20\% for y-polarization. The extracted experimental Q-factors of the resonances show good agreement with the simulated curves. Furthermore, the Q-factor values presented here are comparable to the literature values for THz metasurfaces of asymmetric resonances reported from 20 to 79. \cite{SinghBIC,Fedotov07,AlNaib12,Srivastava15,Gupta16,Al-Naib:15} 

To understand the dependence of incident polarization, the ultra-high asymmetry sample ($\alpha$= 100\%) was measured for angle $\theta$ rotated counter-clockwise from $0^{\circ}$ to $90^{\circ}$ to mimic different linear polarization states; as illustrated in Fig.~3(a).  The resulting spectra for different rotation angles are plotted in Fig.~3(d) with the accompanying simulated spectra,  where $\theta$ in that case corresponds to the polarization angle of the incoming THz wave. The measured Fano resonance disappears when the sample is rotated to $\theta$ = $30^{\circ}$ and reappears at $\theta$ = $45^{\circ}$. The simulated transmission amplitude and Q-factor of the Fano dip are plotted as a function of rotation angle in Fig.~3(b). Furthermore, a colormap for simulated amplitude as a function of incident polarization angle is presented in Fig.~3(c). The transition from Fano to BIC and back to a Fano-type resonance can be observed in the transmission amplitude with a plateau region between $\theta$ = $20^{\circ}$ and $\theta$ = $40^{\circ}$ from both Fig.~3(b) and 3(c). This plateau region is dependent on rotational angles and represents the FW BIC region with vanishing linewidth as described by a dotted circle in pictorial illustration of Fig.~3(c). The Q-factor also becomes infinite as the rotational angles approach the FW BIC region as seen in Fig.~3(b). This is characteristic of a FW BIC arising from destructive interference between radiative modes modulated by a specific parameter. The resonances with high Q-factor around the FW BIC region are termed as supercavity resonances. In previous studies, FW BICs occurred at non-zero incidence angles  \cite{Hsu2013,Yoon2015, Azzam2018} and through geometric manipulation \cite{Han2018,Zhang2018}. The FW BIC experimentally observed in our system is induced by tuning the linear polarization states of electromagnetic field at normal incidence on the sample. This allows for a unique parameter to produce the FW BIC in regions not accessible by symmetry-protected BIC through polarization engineering. The advantage of EM waves is the scalability of optical responses to different frequency bands. Therefore, we can take advantage of this result to induce the polarization selective FW BICs in other deep subwavelength regions by tailoring the metasurface geometry.

To further analyze the behavior of FW BIC in the asymmetric SRR, the transmission spectra from $\theta$ = $0^{\circ}$ to $\theta$ = $90^{\circ}$ is numerically calculated and pictorially illustrated in Fig.~4(a) for different asymmetry $\alpha$ values. For the symmetric sample, there is a flat transmission for all rotational angles as expected. When symmetry is perturbed by $\alpha$ = 20\%, the Fano resonance appears but its linewidth decreases with increasing rotation only to vanish after $\theta$ = $60^{\circ}$. The dipolar resonance above 0.5 THz with low transmission amplitude around 0.2 also starts to split between rotational angles $\theta$ = $40^{\circ}$ and $\theta$ = $60^{\circ}$ for $\alpha \geq$ 40\% as previously reported in four gap circular SRRs \cite{Burrow17}. As the asymmetry is increased, the FW BIC region with vanishing linewidth appears in lower rotational angles with the reappearance of the Fano identifiable at higher angles. As predicted by original FW theory \cite{Freidrich1985}, this occurs due to the change in coupling strength among the resonances that destructively interfere to produce the FW BIC. The dependence of the central rotational degree ($\theta_{c}$) of the FW BIC on asymmetry parameter ($\alpha$) can be fitted exponentially with the equation $\theta_{c}$ = 8.9 + 79.3  exp (-0.013 $\alpha$).
\begin{figure}[t]
\centering\includegraphics[width=\linewidth]{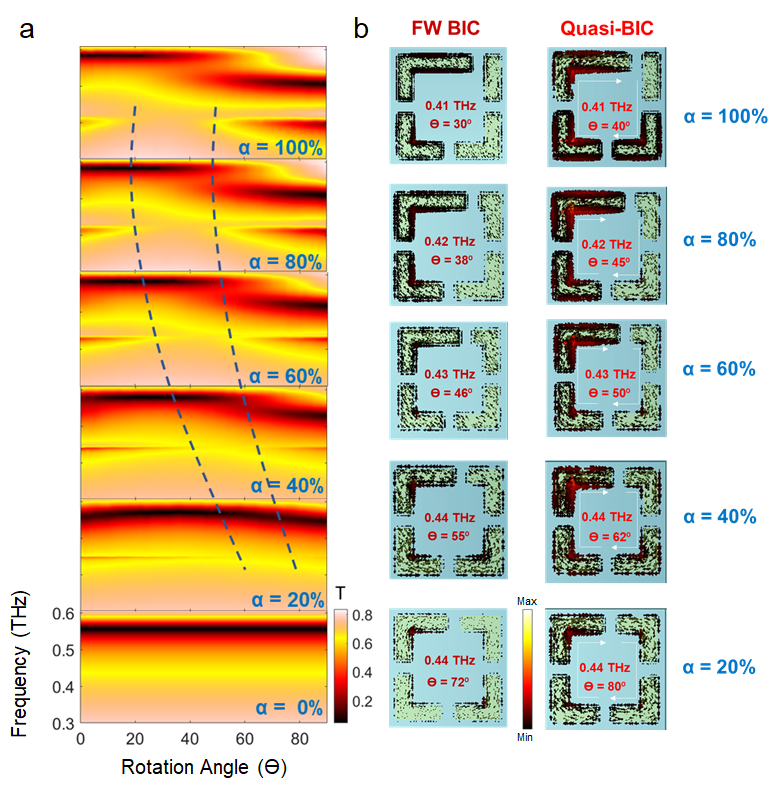}
\caption{(a) Pictorial illustration for the rotational dependence of the transmission amplitude calculated for different asymmetry values $\alpha$. The dotted lines indicate the FW BIC region with very small linewidth. (b) The surface current distributions at rotational angles $\theta$ for BIC and quasi-BIC frequencies for different asymmetry values $\alpha$ indicate the strong coupling to incident radiation as the FW-BIC is perturbed.}
\end{figure}

In addition, the rotational window of FW BIC region changes based on asymmetry as illustrated by dotted lines around the region in Fig.~4(a). Note that the rotational angle corresponding to the linear polarization state of light is directly related to the photonic band dispersion\cite{Shuyuan2019,Doeleman2018}. Further, the Q-factors of the supercavity resonances around the FW BIC region before it diverges to infinity range from 42 to 320 for $\alpha$ = 100\% to $\alpha$ = 20\% respectively. As stated previously, the comparable Q-factors measured from terahertz plasmonic metasurfaces range from 20 to 79 \cite{SinghBIC,Fedotov07,AlNaib12,Srivastava15,Gupta16,Al-Naib:15}. Hence, we have shown the modulation of both the position and Q-factor of the supercavity resonances from FW BIC with degrees of asymmetry. 

To gain an understanding of the physical nature of the FW BIC region, the surface current distributions at the rotational angles for FW BIC and quasi-BIC (angles just to the right of the FW BIC angle) frequencies are plotted for different asymmetry $\alpha$ values in Fig.~4(b). At the FW BIC frequencies, the bound states on the structure are weakly coupled to the radiation as evidenced by low surface current density on the SRR arms in comparison to quasi-BICs. When the FW BIC is perturbed by further rotating the sample to produce the quasi-BICs, the modes within the structure are strongly coupled to the radiation with stronger field confinement on the metallic SRR arms. High current density is concentrated on the top SRR arm which lengthens as the gap is translated. The current distributions exhibit anti-phase current branches as indicated by artificial arrows demonstrating the destructive interference required for FW BIC.

\section{\label{sec:level1}Conclusion\vspace{-1em}}
In summary, we reported experimentally the polarization selective tuning of the Fano resonance into Friedrich-Wintgen bound states in the continuum on an asymmetric terahertz metasurface. The Fano resonance originates from symmetry BICs as they are perturbed by translating a gap of the SRR for horizontal and vertical polarization. The linewidth of the Fano resonance vanishes as a FW BIC manifests at specific linear polarization states. The position and size of this polarization induced FW BIC from the asymmetric sample is calculated to have a strong dependence on the degree of asymmetry given by the asymmetry parameter $\alpha$. High Q-factor supercavity resonances are observed near this FW BIC region of zero linewidth and can be tuned with respect to asymmetry. Modulation of the supercavity resonances allows for an additional degree of freedom in the design of high-Q factor devices. The subwavelength nature of these metasurfaces allows for much smaller mode volume and lifts the dimensional limitations of dielectric photonic crystals. The metadevices studied here have applications in lasing, non-linear optics, and high-performance sensing devices.


\begin{acknowledgments}
This project is supported by the W. M. Keck Foundation, the National Science Foundation (NSF) under NSF award no. 1541959 and 1659224, Air Force Office of Scientific Research (FA9550-16-1-0346) and the NASA Ohio Space Grant (NNX15AL50H).  
C. K. acknowledges support from the Just-Julian Fellowship Program at Howard University and T. A. S. acknowledges support from the CNS Scholars Program. 
\end{acknowledgments}

\bibliography{sample}

\begin{thebibliography}{44}%
\makeatletter
\providecommand \@ifxundefined [1]{%
 \@ifx{#1\undefined}
}%
\providecommand \@ifnum [1]{%
 \ifnum #1\expandafter \@firstoftwo
 \else \expandafter \@secondoftwo
 \fi
}%
\providecommand \@ifx [1]{%
 \ifx #1\expandafter \@firstoftwo
 \else \expandafter \@secondoftwo
 \fi
}%
\providecommand \natexlab [1]{#1}%
\providecommand \enquote  [1]{``#1''}%
\providecommand \bibnamefont  [1]{#1}%
\providecommand \bibfnamefont [1]{#1}%
\providecommand \citenamefont [1]{#1}%
\providecommand \href@noop [0]{\@secondoftwo}%
\providecommand \href [0]{\begingroup \@sanitize@url \@href}%
\providecommand \@href[1]{\@@startlink{#1}\@@href}%
\providecommand \@@href[1]{\endgroup#1\@@endlink}%
\providecommand \@sanitize@url [0]{\catcode `\\12\catcode `\$12\catcode
  `\&12\catcode `\#12\catcode `\^12\catcode `\_12\catcode `\%12\relax}%
\providecommand \@@startlink[1]{}%
\providecommand \@@endlink[0]{}%
\providecommand \url  [0]{\begingroup\@sanitize@url \@url }%
\providecommand \@url [1]{\endgroup\@href {#1}{\urlprefix }}%
\providecommand \urlprefix  [0]{URL }%
\providecommand \Eprint [0]{\href }%
\providecommand \doibase [0]{http://dx.doi.org/}%
\providecommand \selectlanguage [0]{\@gobble}%
\providecommand \bibinfo  [0]{\@secondoftwo}%
\providecommand \bibfield  [0]{\@secondoftwo}%
\providecommand \translation [1]{[#1]}%
\providecommand \BibitemOpen [0]{}%
\providecommand \bibitemStop [0]{}%
\providecommand \bibitemNoStop [0]{.\EOS\space}%
\providecommand \EOS [0]{\spacefactor3000\relax}%
\providecommand \BibitemShut  [1]{\csname bibitem#1\endcsname}%
\let\auto@bib@innerbib\@empty
\bibitem [{\citenamefont {{von Neuman}}\ and\ \citenamefont
  {{Wigner}}(1929)}]{von1929}%
  \BibitemOpen
  \bibfield  {author} {\bibinfo {author} {\bibfnamefont {J.}~\bibnamefont {{von
  Neuman}}}\ and\ \bibinfo {author} {\bibfnamefont {E.}~\bibnamefont
  {{Wigner}}},\ }\href@noop {} {\bibfield  {journal} {\bibinfo  {journal}
  {Physikalische Zeitschrift}\ }\textbf {\bibinfo {volume} {30}},\ \bibinfo
  {pages} {467} (\bibinfo {year} {1929})}\BibitemShut {NoStop}%
\bibitem [{\citenamefont {Schult}\ \emph {et~al.}(1989)\citenamefont {Schult},
  \citenamefont {Ravenhall},\ and\ \citenamefont {Wyld}}]{Schult89}%
  \BibitemOpen
  \bibfield  {author} {\bibinfo {author} {\bibfnamefont {R.~L.}\ \bibnamefont
  {Schult}}, \bibinfo {author} {\bibfnamefont {D.~G.}\ \bibnamefont
  {Ravenhall}}, \ and\ \bibinfo {author} {\bibfnamefont {H.~W.}\ \bibnamefont
  {Wyld}},\ }\href {\doibase 10.1103/PhysRevB.39.5476} {\bibfield  {journal}
  {\bibinfo  {journal} {Phys. Rev. B}\ }\textbf {\bibinfo {volume} {39}},\
  \bibinfo {pages} {5476} (\bibinfo {year} {1989})}\BibitemShut {NoStop}%
\bibitem [{\citenamefont {Exner}\ \emph {et~al.}(1996)\citenamefont {Exner},
  \citenamefont {Šeba}, \citenamefont {Tater},\ and\ \citenamefont
  {Vaněk}}]{Exner1996}%
  \BibitemOpen
  \bibfield  {author} {\bibinfo {author} {\bibfnamefont {P.}~\bibnamefont
  {Exner}}, \bibinfo {author} {\bibfnamefont {P.}~\bibnamefont {Šeba}},
  \bibinfo {author} {\bibfnamefont {M.}~\bibnamefont {Tater}}, \ and\ \bibinfo
  {author} {\bibfnamefont {D.}~\bibnamefont {Vaněk}},\ }\href {\doibase
  10.1063/1.531673} {\bibfield  {journal} {\bibinfo  {journal} {Journal of
  Mathematical Physics}\ }\textbf {\bibinfo {volume} {37}},\ \bibinfo {pages}
  {4867} (\bibinfo {year} {1996})}\BibitemShut {NoStop}%
\bibitem [{\citenamefont {Parker}(1966)}]{Parker66}%
  \BibitemOpen
  \bibfield  {author} {\bibinfo {author} {\bibfnamefont {R.}~\bibnamefont
  {Parker}},\ }\href {\doibase 10.1016/0022-460X(66)90154-4} {\bibfield
  {journal} {\bibinfo  {journal} {Journal of Sound and Vibration}\ }\textbf
  {\bibinfo {volume} {4}},\ \bibinfo {pages} {62} (\bibinfo {year}
  {1966})}\BibitemShut {NoStop}%
\bibitem [{\citenamefont {Ursell}(1951)}]{ursell51}%
  \BibitemOpen
  \bibfield  {author} {\bibinfo {author} {\bibfnamefont {F.}~\bibnamefont
  {Ursell}},\ }\href {\doibase 10.1017/S0305004100026700} {\bibfield  {journal}
  {\bibinfo  {journal} {Mathematical Proceedings of the Cambridge Philosophical
  Society}\ }\textbf {\bibinfo {volume} {47}},\ \bibinfo {pages} {347}
  (\bibinfo {year} {1951})}\BibitemShut {NoStop}%
\bibitem [{\citenamefont {Watts}\ \emph {et~al.}(2002)\citenamefont {Watts},
  \citenamefont {Johnson}, \citenamefont {Haus},\ and\ \citenamefont
  {Joannopoulos}}]{Watts:02}%
  \BibitemOpen
  \bibfield  {author} {\bibinfo {author} {\bibfnamefont {M.~R.}\ \bibnamefont
  {Watts}}, \bibinfo {author} {\bibfnamefont {S.~G.}\ \bibnamefont {Johnson}},
  \bibinfo {author} {\bibfnamefont {H.~A.}\ \bibnamefont {Haus}}, \ and\
  \bibinfo {author} {\bibfnamefont {J.~D.}\ \bibnamefont {Joannopoulos}},\
  }\href {\doibase 10.1364/OL.27.001785} {\bibfield  {journal} {\bibinfo
  {journal} {Opt. Lett.}\ }\textbf {\bibinfo {volume} {27}},\ \bibinfo {pages}
  {1785} (\bibinfo {year} {2002})}\BibitemShut {NoStop}%
\bibitem [{\citenamefont {Marinica}\ \emph {et~al.}(2008)\citenamefont
  {Marinica}, \citenamefont {Borisov},\ and\ \citenamefont
  {Shabanov}}]{Marinica08}%
  \BibitemOpen
  \bibfield  {author} {\bibinfo {author} {\bibfnamefont {D.~C.}\ \bibnamefont
  {Marinica}}, \bibinfo {author} {\bibfnamefont {A.~G.}\ \bibnamefont
  {Borisov}}, \ and\ \bibinfo {author} {\bibfnamefont {S.~V.}\ \bibnamefont
  {Shabanov}},\ }\href {\doibase 10.1103/PhysRevLett.100.183902} {\bibfield
  {journal} {\bibinfo  {journal} {Phys. Rev. Lett.}\ }\textbf {\bibinfo
  {volume} {100}},\ \bibinfo {pages} {183902} (\bibinfo {year}
  {2008})}\BibitemShut {NoStop}%
\bibitem [{\citenamefont {Molina}\ \emph {et~al.}(2012)\citenamefont {Molina},
  \citenamefont {Miroshnichenko},\ and\ \citenamefont {Kivshar}}]{Molina12}%
  \BibitemOpen
  \bibfield  {author} {\bibinfo {author} {\bibfnamefont {M.~I.}\ \bibnamefont
  {Molina}}, \bibinfo {author} {\bibfnamefont {A.~E.}\ \bibnamefont
  {Miroshnichenko}}, \ and\ \bibinfo {author} {\bibfnamefont {Y.~S.}\
  \bibnamefont {Kivshar}},\ }\href {\doibase 10.1103/PhysRevLett.108.070401}
  {\bibfield  {journal} {\bibinfo  {journal} {Phys. Rev. Lett.}\ }\textbf
  {\bibinfo {volume} {108}},\ \bibinfo {pages} {070401} (\bibinfo {year}
  {2012})}\BibitemShut {NoStop}%
\bibitem [{\citenamefont {Monticone}\ and\ \citenamefont
  {Al\`u}(2014)}]{Monticone14}%
  \BibitemOpen
  \bibfield  {author} {\bibinfo {author} {\bibfnamefont {F.}~\bibnamefont
  {Monticone}}\ and\ \bibinfo {author} {\bibfnamefont {A.}~\bibnamefont
  {Al\`u}},\ }\href {\doibase 10.1103/PhysRevLett.112.213903} {\bibfield
  {journal} {\bibinfo  {journal} {Phys. Rev. Lett.}\ }\textbf {\bibinfo
  {volume} {112}},\ \bibinfo {pages} {213903} (\bibinfo {year}
  {2014})}\BibitemShut {NoStop}%
\bibitem [{\citenamefont {Hsu}\ \emph {et~al.}(2013)\citenamefont {Hsu},
  \citenamefont {Zhen}, \citenamefont {Lee}, \citenamefont {Chua},
  \citenamefont {Johnson}, \citenamefont {Joannopoulos},\ and\ \citenamefont
  {Solja{\v{c}}i{\'{c}}}}]{Hsu2013}%
  \BibitemOpen
  \bibfield  {author} {\bibinfo {author} {\bibfnamefont {C.~W.}\ \bibnamefont
  {Hsu}}, \bibinfo {author} {\bibfnamefont {B.}~\bibnamefont {Zhen}}, \bibinfo
  {author} {\bibfnamefont {J.}~\bibnamefont {Lee}}, \bibinfo {author}
  {\bibfnamefont {S.-L.}\ \bibnamefont {Chua}}, \bibinfo {author}
  {\bibfnamefont {S.~G.}\ \bibnamefont {Johnson}}, \bibinfo {author}
  {\bibfnamefont {J.~D.}\ \bibnamefont {Joannopoulos}}, \ and\ \bibinfo
  {author} {\bibfnamefont {M.}~\bibnamefont {Solja{\v{c}}i{\'{c}}}},\ }\href
  {\doibase 10.1038/nature12289} {\bibfield  {journal} {\bibinfo  {journal}
  {Nature}\ }\textbf {\bibinfo {volume} {499}},\ \bibinfo {pages} {188}
  (\bibinfo {year} {2013})}\BibitemShut {NoStop}%
\bibitem [{\citenamefont {Jin}\ \emph {et~al.}(2019)\citenamefont {Jin},
  \citenamefont {Yin}, \citenamefont {Ni}, \citenamefont {Soljačić},
  \citenamefont {Zhen},\ and\ \citenamefont {Peng}}]{Jin2019}%
  \BibitemOpen
  \bibfield  {author} {\bibinfo {author} {\bibfnamefont {J.}~\bibnamefont
  {Jin}}, \bibinfo {author} {\bibfnamefont {X.}~\bibnamefont {Yin}}, \bibinfo
  {author} {\bibfnamefont {L.}~\bibnamefont {Ni}}, \bibinfo {author}
  {\bibfnamefont {M.}~\bibnamefont {Soljačić}}, \bibinfo {author}
  {\bibfnamefont {B.}~\bibnamefont {Zhen}}, \ and\ \bibinfo {author}
  {\bibfnamefont {C.}~\bibnamefont {Peng}},\ }\href {\doibase
  10.1038/s41586-019-1664-7} {\bibfield  {journal} {\bibinfo  {journal}
  {Nature}\ }\textbf {\bibinfo {volume} {574}},\ \bibinfo {pages} {501}
  (\bibinfo {year} {2019})}\BibitemShut {NoStop}%
\bibitem [{\citenamefont {Li}\ \emph {et~al.}(2019)\citenamefont {Li},
  \citenamefont {Zhou}, \citenamefont {Liu},\ and\ \citenamefont
  {Xiao}}]{Shuyuan2019}%
  \BibitemOpen
  \bibfield  {author} {\bibinfo {author} {\bibfnamefont {S.}~\bibnamefont
  {Li}}, \bibinfo {author} {\bibfnamefont {C.}~\bibnamefont {Zhou}}, \bibinfo
  {author} {\bibfnamefont {T.}~\bibnamefont {Liu}}, \ and\ \bibinfo {author}
  {\bibfnamefont {S.}~\bibnamefont {Xiao}},\ }\href {\doibase
  10.1103/PhysRevA.100.063803} {\bibfield  {journal} {\bibinfo  {journal}
  {Phys. Rev. A}\ }\textbf {\bibinfo {volume} {100}},\ \bibinfo {pages}
  {063803} (\bibinfo {year} {2019})}\BibitemShut {NoStop}%
\bibitem [{\citenamefont {Rybin}\ and\ \citenamefont
  {Kivshar}(2017)}]{Rybin2017}%
  \BibitemOpen
  \bibfield  {author} {\bibinfo {author} {\bibfnamefont {M.}~\bibnamefont
  {Rybin}}\ and\ \bibinfo {author} {\bibfnamefont {Y.}~\bibnamefont
  {Kivshar}},\ }\href {\doibase 10.1038/541164a} {\bibfield  {journal}
  {\bibinfo  {journal} {Nature}\ }\textbf {\bibinfo {volume} {541}},\ \bibinfo
  {pages} {164} (\bibinfo {year} {2017})}\BibitemShut {NoStop}%
\bibitem [{\citenamefont {Ha}\ \emph {et~al.}(2018)\citenamefont {Ha},
  \citenamefont {Fu}, \citenamefont {Emani}, \citenamefont {Pan}, \citenamefont
  {Bakker}, \citenamefont {Paniagua-Dom{\'{\i}}nguez},\ and\ \citenamefont
  {Kuznetsov}}]{Ha2018}%
  \BibitemOpen
  \bibfield  {author} {\bibinfo {author} {\bibfnamefont {S.~T.}\ \bibnamefont
  {Ha}}, \bibinfo {author} {\bibfnamefont {Y.~H.}\ \bibnamefont {Fu}}, \bibinfo
  {author} {\bibfnamefont {N.~K.}\ \bibnamefont {Emani}}, \bibinfo {author}
  {\bibfnamefont {Z.}~\bibnamefont {Pan}}, \bibinfo {author} {\bibfnamefont
  {R.~M.}\ \bibnamefont {Bakker}}, \bibinfo {author} {\bibfnamefont
  {R.}~\bibnamefont {Paniagua-Dom{\'{\i}}nguez}}, \ and\ \bibinfo {author}
  {\bibfnamefont {A.~I.}\ \bibnamefont {Kuznetsov}},\ }\href {\doibase
  10.1038/s41565-018-0245-5} {\bibfield  {journal} {\bibinfo  {journal} {Nature
  Nanotechnology}\ }\textbf {\bibinfo {volume} {13}},\ \bibinfo {pages} {1042}
  (\bibinfo {year} {2018})}\BibitemShut {NoStop}%
\bibitem [{\citenamefont {Kodigala}\ \emph {et~al.}(2017)\citenamefont
  {Kodigala}, \citenamefont {Lepetit}, \citenamefont {Gu}, \citenamefont
  {Bahari}, \citenamefont {Fainman},\ and\ \citenamefont
  {Kant{\'e}}}]{Kodigala2017}%
  \BibitemOpen
  \bibfield  {author} {\bibinfo {author} {\bibfnamefont {A.}~\bibnamefont
  {Kodigala}}, \bibinfo {author} {\bibfnamefont {T.}~\bibnamefont {Lepetit}},
  \bibinfo {author} {\bibfnamefont {Q.}~\bibnamefont {Gu}}, \bibinfo {author}
  {\bibfnamefont {B.}~\bibnamefont {Bahari}}, \bibinfo {author} {\bibfnamefont
  {Y.}~\bibnamefont {Fainman}}, \ and\ \bibinfo {author} {\bibfnamefont
  {B.}~\bibnamefont {Kant{\'e}}},\ }\href {https://doi.org/10.1038/nature20799}
  {\bibfield  {journal} {\bibinfo  {journal} {Nature}\ }\textbf {\bibinfo
  {volume} {541}},\ \bibinfo {pages} {196 EP } (\bibinfo {year}
  {2017})}\BibitemShut {NoStop}%
\bibitem [{\citenamefont {Liu}\ \emph {et~al.}(2012)\citenamefont {Liu},
  \citenamefont {Hwang}, \citenamefont {Tao}, \citenamefont {Strikwerda},
  \citenamefont {Fan}, \citenamefont {Keiser}, \citenamefont {Sternbach},
  \citenamefont {West}, \citenamefont {Kittiwatanakul}, \citenamefont {Lu},
  \citenamefont {Wolf}, \citenamefont {Omenetto}, \citenamefont {Zhang},
  \citenamefont {Nelson},\ and\ \citenamefont {Averitt}}]{Liu2012}%
  \BibitemOpen
  \bibfield  {author} {\bibinfo {author} {\bibfnamefont {M.}~\bibnamefont
  {Liu}}, \bibinfo {author} {\bibfnamefont {H.~Y.}\ \bibnamefont {Hwang}},
  \bibinfo {author} {\bibfnamefont {H.}~\bibnamefont {Tao}}, \bibinfo {author}
  {\bibfnamefont {A.~C.}\ \bibnamefont {Strikwerda}}, \bibinfo {author}
  {\bibfnamefont {K.}~\bibnamefont {Fan}}, \bibinfo {author} {\bibfnamefont
  {G.~R.}\ \bibnamefont {Keiser}}, \bibinfo {author} {\bibfnamefont {A.~J.}\
  \bibnamefont {Sternbach}}, \bibinfo {author} {\bibfnamefont {K.~G.}\
  \bibnamefont {West}}, \bibinfo {author} {\bibfnamefont {S.}~\bibnamefont
  {Kittiwatanakul}}, \bibinfo {author} {\bibfnamefont {J.}~\bibnamefont {Lu}},
  \bibinfo {author} {\bibfnamefont {S.~A.}\ \bibnamefont {Wolf}}, \bibinfo
  {author} {\bibfnamefont {F.~G.}\ \bibnamefont {Omenetto}}, \bibinfo {author}
  {\bibfnamefont {X.}~\bibnamefont {Zhang}}, \bibinfo {author} {\bibfnamefont
  {K.~A.}\ \bibnamefont {Nelson}}, \ and\ \bibinfo {author} {\bibfnamefont
  {R.~D.}\ \bibnamefont {Averitt}},\ }\href {\doibase 10.1038/nature11231}
  {\bibfield  {journal} {\bibinfo  {journal} {Nature}\ }\textbf {\bibinfo
  {volume} {487}},\ \bibinfo {pages} {345} (\bibinfo {year}
  {2012})}\BibitemShut {NoStop}%
\bibitem [{\citenamefont {Koshelev}\ \emph
  {et~al.}(2018{\natexlab{a}})\citenamefont {Koshelev}, \citenamefont
  {Bogdanov},\ and\ \citenamefont {Kivshar}}]{Koshelev2018}%
  \BibitemOpen
  \bibfield  {author} {\bibinfo {author} {\bibfnamefont {K.}~\bibnamefont
  {Koshelev}}, \bibinfo {author} {\bibfnamefont {A.}~\bibnamefont {Bogdanov}},
  \ and\ \bibinfo {author} {\bibfnamefont {Y.}~\bibnamefont {Kivshar}},\ }\href
  {\doibase 10.1016/j.scib.2018.12.003} {\bibfield  {journal} {\bibinfo
  {journal} {Science Bulletin}\ } (\bibinfo {year} {2018}{\natexlab{a}}),\
  10.1016/j.scib.2018.12.003}\BibitemShut {NoStop}%
\bibitem [{\citenamefont {Vabishchevich}\ \emph {et~al.}(2018)\citenamefont
  {Vabishchevich}, \citenamefont {Liu}, \citenamefont {Sinclair}, \citenamefont
  {Keeler}, \citenamefont {Peake},\ and\ \citenamefont
  {Brener}}]{Vabishchevich2018}%
  \BibitemOpen
  \bibfield  {author} {\bibinfo {author} {\bibfnamefont {P.~P.}\ \bibnamefont
  {Vabishchevich}}, \bibinfo {author} {\bibfnamefont {S.}~\bibnamefont {Liu}},
  \bibinfo {author} {\bibfnamefont {M.~B.}\ \bibnamefont {Sinclair}}, \bibinfo
  {author} {\bibfnamefont {G.~A.}\ \bibnamefont {Keeler}}, \bibinfo {author}
  {\bibfnamefont {G.~M.}\ \bibnamefont {Peake}}, \ and\ \bibinfo {author}
  {\bibfnamefont {I.}~\bibnamefont {Brener}},\ }\href {\doibase
  10.1021/acsphotonics.7b01478} {\bibfield  {journal} {\bibinfo  {journal}
  {{ACS} Photonics}\ }\textbf {\bibinfo {volume} {5}},\ \bibinfo {pages} {1685}
  (\bibinfo {year} {2018})}\BibitemShut {NoStop}%
\bibitem [{\citenamefont {Yahiaoui}\ \emph {et~al.}(2015)\citenamefont
  {Yahiaoui}, \citenamefont {Tan}, \citenamefont {Cong}, \citenamefont {Singh},
  \citenamefont {Yan},\ and\ \citenamefont {Zhang}}]{Yahiaoui2015}%
  \BibitemOpen
  \bibfield  {author} {\bibinfo {author} {\bibfnamefont {R.}~\bibnamefont
  {Yahiaoui}}, \bibinfo {author} {\bibfnamefont {S.}~\bibnamefont {Tan}},
  \bibinfo {author} {\bibfnamefont {L.}~\bibnamefont {Cong}}, \bibinfo {author}
  {\bibfnamefont {R.}~\bibnamefont {Singh}}, \bibinfo {author} {\bibfnamefont
  {F.}~\bibnamefont {Yan}}, \ and\ \bibinfo {author} {\bibfnamefont
  {W.}~\bibnamefont {Zhang}},\ }\href {\doibase 10.1063/1.4929449} {\bibfield
  {journal} {\bibinfo  {journal} {Journal of Applied Physics}\ }\textbf
  {\bibinfo {volume} {118}},\ \bibinfo {pages} {083103} (\bibinfo {year}
  {2015})}\BibitemShut {NoStop}%
\bibitem [{\citenamefont {Evlyukhin}\ \emph {et~al.}(2010)\citenamefont
  {Evlyukhin}, \citenamefont {Bozhevolnyi}, \citenamefont {Pors}, \citenamefont
  {Nielsen}, \citenamefont {Radko}, \citenamefont {Willatzen},\ and\
  \citenamefont {Albrektsen}}]{Evlyukhin2010}%
  \BibitemOpen
  \bibfield  {author} {\bibinfo {author} {\bibfnamefont {A.~B.}\ \bibnamefont
  {Evlyukhin}}, \bibinfo {author} {\bibfnamefont {S.~I.}\ \bibnamefont
  {Bozhevolnyi}}, \bibinfo {author} {\bibfnamefont {A.}~\bibnamefont {Pors}},
  \bibinfo {author} {\bibfnamefont {M.~G.}\ \bibnamefont {Nielsen}}, \bibinfo
  {author} {\bibfnamefont {I.~P.}\ \bibnamefont {Radko}}, \bibinfo {author}
  {\bibfnamefont {M.}~\bibnamefont {Willatzen}}, \ and\ \bibinfo {author}
  {\bibfnamefont {O.}~\bibnamefont {Albrektsen}},\ }\href {\doibase
  10.1021/nl102572q} {\bibfield  {journal} {\bibinfo  {journal} {Nano Letters}\
  }\textbf {\bibinfo {volume} {10}},\ \bibinfo {pages} {4571} (\bibinfo {year}
  {2010})}\BibitemShut {NoStop}%
\bibitem [{\citenamefont {Liu}\ \emph {et~al.}(2017)\citenamefont {Liu},
  \citenamefont {Wang}, \citenamefont {Zhao}, \citenamefont {Zhou},\ and\
  \citenamefont {Sun}}]{Liu:17}%
  \BibitemOpen
  \bibfield  {author} {\bibinfo {author} {\bibfnamefont {Y.}~\bibnamefont
  {Liu}}, \bibinfo {author} {\bibfnamefont {S.}~\bibnamefont {Wang}}, \bibinfo
  {author} {\bibfnamefont {D.}~\bibnamefont {Zhao}}, \bibinfo {author}
  {\bibfnamefont {W.}~\bibnamefont {Zhou}}, \ and\ \bibinfo {author}
  {\bibfnamefont {Y.}~\bibnamefont {Sun}},\ }\href {\doibase
  10.1364/OE.25.010536} {\bibfield  {journal} {\bibinfo  {journal} {Opt.
  Express}\ }\textbf {\bibinfo {volume} {25}},\ \bibinfo {pages} {10536}
  (\bibinfo {year} {2017})}\BibitemShut {NoStop}%
\bibitem [{\citenamefont {Friedrich}\ and\ \citenamefont
  {Wintgen}(1985)}]{Freidrich1985}%
  \BibitemOpen
  \bibfield  {author} {\bibinfo {author} {\bibfnamefont {H.}~\bibnamefont
  {Friedrich}}\ and\ \bibinfo {author} {\bibfnamefont {D.}~\bibnamefont
  {Wintgen}},\ }\href {\doibase 10.1103/PhysRevA.32.3231} {\bibfield  {journal}
  {\bibinfo  {journal} {Phys. Rev. A}\ }\textbf {\bibinfo {volume} {32}},\
  \bibinfo {pages} {3231} (\bibinfo {year} {1985})}\BibitemShut {NoStop}%
\bibitem [{\citenamefont {Azzam}\ \emph {et~al.}(2018)\citenamefont {Azzam},
  \citenamefont {Shalaev}, \citenamefont {Boltasseva},\ and\ \citenamefont
  {Kildishev}}]{Azzam2018}%
  \BibitemOpen
  \bibfield  {author} {\bibinfo {author} {\bibfnamefont {S.~I.}\ \bibnamefont
  {Azzam}}, \bibinfo {author} {\bibfnamefont {V.~M.}\ \bibnamefont {Shalaev}},
  \bibinfo {author} {\bibfnamefont {A.}~\bibnamefont {Boltasseva}}, \ and\
  \bibinfo {author} {\bibfnamefont {A.~V.}\ \bibnamefont {Kildishev}},\ }\href
  {\doibase 10.1103/PhysRevLett.121.253901} {\bibfield  {journal} {\bibinfo
  {journal} {Phys. Rev. Lett.}\ }\textbf {\bibinfo {volume} {121}},\ \bibinfo
  {pages} {253901} (\bibinfo {year} {2018})}\BibitemShut {NoStop}%
\bibitem [{\citenamefont {Han}\ \emph {et~al.}(2019)\citenamefont {Han},
  \citenamefont {Cong}, \citenamefont {Srivastava}, \citenamefont {Qiang},
  \citenamefont {Rybin}, \citenamefont {Kumar}, \citenamefont {Jain},
  \citenamefont {Lim}, \citenamefont {Achanta}, \citenamefont {Prabhu},
  \citenamefont {Wang}, \citenamefont {Kivshar},\ and\ \citenamefont
  {Singh}}]{Han2018}%
  \BibitemOpen
  \bibfield  {author} {\bibinfo {author} {\bibfnamefont {S.}~\bibnamefont
  {Han}}, \bibinfo {author} {\bibfnamefont {L.}~\bibnamefont {Cong}}, \bibinfo
  {author} {\bibfnamefont {Y.~K.}\ \bibnamefont {Srivastava}}, \bibinfo
  {author} {\bibfnamefont {B.}~\bibnamefont {Qiang}}, \bibinfo {author}
  {\bibfnamefont {M.~V.}\ \bibnamefont {Rybin}}, \bibinfo {author}
  {\bibfnamefont {A.}~\bibnamefont {Kumar}}, \bibinfo {author} {\bibfnamefont
  {R.}~\bibnamefont {Jain}}, \bibinfo {author} {\bibfnamefont {W.~X.}\
  \bibnamefont {Lim}}, \bibinfo {author} {\bibfnamefont {V.~G.}\ \bibnamefont
  {Achanta}}, \bibinfo {author} {\bibfnamefont {S.~S.}\ \bibnamefont {Prabhu}},
  \bibinfo {author} {\bibfnamefont {Q.~J.}\ \bibnamefont {Wang}}, \bibinfo
  {author} {\bibfnamefont {Y.~S.}\ \bibnamefont {Kivshar}}, \ and\ \bibinfo
  {author} {\bibfnamefont {R.}~\bibnamefont {Singh}},\ }\href {\doibase
  10.1002/adma.201901921} {\bibfield  {journal} {\bibinfo  {journal} {Advanced
  Materials}\ }\textbf {\bibinfo {volume} {31}},\ \bibinfo {pages} {1901921}
  (\bibinfo {year} {2019})}\BibitemShut {NoStop}%
\bibitem [{\citenamefont {Kikkawa}\ \emph {et~al.}(2019)\citenamefont
  {Kikkawa}, \citenamefont {Nishida},\ and\ \citenamefont
  {Kadoya}}]{Kikkawa2019}%
  \BibitemOpen
  \bibfield  {author} {\bibinfo {author} {\bibfnamefont {R.}~\bibnamefont
  {Kikkawa}}, \bibinfo {author} {\bibfnamefont {M.}~\bibnamefont {Nishida}}, \
  and\ \bibinfo {author} {\bibfnamefont {Y.}~\bibnamefont {Kadoya}},\ }\href
  {\doibase 10.1088/1367-2630/ab4f54} {\bibfield  {journal} {\bibinfo
  {journal} {New Journal of Physics}\ }\textbf {\bibinfo {volume} {21}},\
  \bibinfo {pages} {113020} (\bibinfo {year} {2019})}\BibitemShut {NoStop}%
\bibitem [{\citenamefont {Feshbach}(1958)}]{Feshbach1958}%
  \BibitemOpen
  \bibfield  {author} {\bibinfo {author} {\bibfnamefont {H.}~\bibnamefont
  {Feshbach}},\ }\href {\doibase https://doi.org/10.1016/0003-4916(58)90007-1}
  {\bibfield  {journal} {\bibinfo  {journal} {Annals of Physics}\ }\textbf
  {\bibinfo {volume} {5}},\ \bibinfo {pages} {357 } (\bibinfo {year}
  {1958})}\BibitemShut {NoStop}%
\bibitem [{\citenamefont {Yoon}\ \emph {et~al.}(2015)\citenamefont {Yoon},
  \citenamefont {Song},\ and\ \citenamefont {Magnusson}}]{Yoon2015}%
  \BibitemOpen
  \bibfield  {author} {\bibinfo {author} {\bibfnamefont {J.~W.}\ \bibnamefont
  {Yoon}}, \bibinfo {author} {\bibfnamefont {S.~H.}\ \bibnamefont {Song}}, \
  and\ \bibinfo {author} {\bibfnamefont {R.}~\bibnamefont {Magnusson}},\ }\href
  {https://doi.org/10.1038/srep18301} {\bibfield  {journal} {\bibinfo
  {journal} {Scientific Reports}\ }\textbf {\bibinfo {volume} {5}},\ \bibinfo
  {pages} {18301 EP } (\bibinfo {year} {2015})},\ \bibinfo {note}
  {article}\BibitemShut {NoStop}%
\bibitem [{\citenamefont {Cong}\ and\ \citenamefont {Singh}(2019)}]{SinghBIC}%
  \BibitemOpen
  \bibfield  {author} {\bibinfo {author} {\bibfnamefont {L.}~\bibnamefont
  {Cong}}\ and\ \bibinfo {author} {\bibfnamefont {R.}~\bibnamefont {Singh}},\
  }\href {\doibase 10.1002/adom.201900383} {\bibfield  {journal} {\bibinfo
  {journal} {Advanced Optical Materials}\ }\textbf {\bibinfo {volume} {7}},\
  \bibinfo {pages} {1900383} (\bibinfo {year} {2019})}\BibitemShut {NoStop}%
\bibitem [{\citenamefont {Abujetas}\ \emph {et~al.}(2019)\citenamefont
  {Abujetas}, \citenamefont {van Hoof}, \citenamefont {ter Huurne},
  \citenamefont {Rivas},\ and\ \citenamefont {S\'{a}nchez-Gil}}]{Abujetas2019}%
  \BibitemOpen
  \bibfield  {author} {\bibinfo {author} {\bibfnamefont {D.~R.}\ \bibnamefont
  {Abujetas}}, \bibinfo {author} {\bibfnamefont {N.}~\bibnamefont {van Hoof}},
  \bibinfo {author} {\bibfnamefont {S.}~\bibnamefont {ter Huurne}}, \bibinfo
  {author} {\bibfnamefont {J.~G.}\ \bibnamefont {Rivas}}, \ and\ \bibinfo
  {author} {\bibfnamefont {J.~A.}\ \bibnamefont {S\'{a}nchez-Gil}},\ }\href
  {\doibase 10.1364/OPTICA.6.000996} {\bibfield  {journal} {\bibinfo  {journal}
  {Optica}\ }\textbf {\bibinfo {volume} {6}},\ \bibinfo {pages} {996} (\bibinfo
  {year} {2019})}\BibitemShut {NoStop}%
\bibitem [{\citenamefont {Fan}\ \emph {et~al.}(2019)\citenamefont {Fan},
  \citenamefont {Shadrivov},\ and\ \citenamefont {Padilla}}]{Fan2019}%
  \BibitemOpen
  \bibfield  {author} {\bibinfo {author} {\bibfnamefont {K.}~\bibnamefont
  {Fan}}, \bibinfo {author} {\bibfnamefont {I.~V.}\ \bibnamefont {Shadrivov}},
  \ and\ \bibinfo {author} {\bibfnamefont {W.~J.}\ \bibnamefont {Padilla}},\
  }\href {\doibase 10.1364/OPTICA.6.000169} {\bibfield  {journal} {\bibinfo
  {journal} {Optica}\ }\textbf {\bibinfo {volume} {6}},\ \bibinfo {pages} {169}
  (\bibinfo {year} {2019})}\BibitemShut {NoStop}%
\bibitem [{\citenamefont {Zhang}\ \emph {et~al.}(2018)\citenamefont {Zhang},
  \citenamefont {Charous}, \citenamefont {Nagai}, \citenamefont {Mittleman},\
  and\ \citenamefont {Mendis}}]{Zhang2018}%
  \BibitemOpen
  \bibfield  {author} {\bibinfo {author} {\bibfnamefont {W.}~\bibnamefont
  {Zhang}}, \bibinfo {author} {\bibfnamefont {A.}~\bibnamefont {Charous}},
  \bibinfo {author} {\bibfnamefont {M.}~\bibnamefont {Nagai}}, \bibinfo
  {author} {\bibfnamefont {D.~M.}\ \bibnamefont {Mittleman}}, \ and\ \bibinfo
  {author} {\bibfnamefont {R.}~\bibnamefont {Mendis}},\ }\href {\doibase
  10.1364/OE.26.013195} {\bibfield  {journal} {\bibinfo  {journal} {Opt.
  Express}\ }\textbf {\bibinfo {volume} {26}},\ \bibinfo {pages} {13195}
  (\bibinfo {year} {2018})}\BibitemShut {NoStop}%
\bibitem [{\citenamefont {Burrow}\ \emph {et~al.}(2017)\citenamefont {Burrow},
  \citenamefont {Yahiaoui}, \citenamefont {Sarangan}, \citenamefont {Agha},
  \citenamefont {Mathews},\ and\ \citenamefont {Searles}}]{Burrow17}%
  \BibitemOpen
  \bibfield  {author} {\bibinfo {author} {\bibfnamefont {J.~A.}\ \bibnamefont
  {Burrow}}, \bibinfo {author} {\bibfnamefont {R.}~\bibnamefont {Yahiaoui}},
  \bibinfo {author} {\bibfnamefont {A.}~\bibnamefont {Sarangan}}, \bibinfo
  {author} {\bibfnamefont {I.}~\bibnamefont {Agha}}, \bibinfo {author}
  {\bibfnamefont {J.}~\bibnamefont {Mathews}}, \ and\ \bibinfo {author}
  {\bibfnamefont {T.~A.}\ \bibnamefont {Searles}},\ }\href@noop {} {\bibfield
  {journal} {\bibinfo  {journal} {Opt. Express}\ }\textbf {\bibinfo {volume}
  {25}},\ \bibinfo {pages} {32540} (\bibinfo {year} {2017})}\BibitemShut
  {NoStop}%
\bibitem [{\citenamefont {Singh}\ \emph
  {et~al.}(2011{\natexlab{a}})\citenamefont {Singh}, \citenamefont {Al-Naib},
  \citenamefont {Koch},\ and\ \citenamefont {Zhang}}]{Singh:11}%
  \BibitemOpen
  \bibfield  {author} {\bibinfo {author} {\bibfnamefont {R.}~\bibnamefont
  {Singh}}, \bibinfo {author} {\bibfnamefont {I.}~\bibnamefont {Al-Naib}},
  \bibinfo {author} {\bibfnamefont {M.}~\bibnamefont {Koch}}, \ and\ \bibinfo
  {author} {\bibfnamefont {W.}~\bibnamefont {Zhang}},\ }\href@noop {}
  {\bibfield  {journal} {\bibinfo  {journal} {Opt. Express}\ }\textbf {\bibinfo
  {volume} {19}},\ \bibinfo {pages} {6312} (\bibinfo {year}
  {2011}{\natexlab{a}})}\BibitemShut {NoStop}%
\bibitem [{\citenamefont {Cong}\ \emph {et~al.}(2015)\citenamefont {Cong},
  \citenamefont {Manjappa}, \citenamefont {Xu}, \citenamefont {Al-Naib},
  \citenamefont {Zhang},\ and\ \citenamefont {Singh}}]{Cong:15}%
  \BibitemOpen
  \bibfield  {author} {\bibinfo {author} {\bibfnamefont {L.}~\bibnamefont
  {Cong}}, \bibinfo {author} {\bibfnamefont {M.}~\bibnamefont {Manjappa}},
  \bibinfo {author} {\bibfnamefont {N.}~\bibnamefont {Xu}}, \bibinfo {author}
  {\bibfnamefont {I.}~\bibnamefont {Al-Naib}}, \bibinfo {author} {\bibfnamefont
  {W.}~\bibnamefont {Zhang}}, \ and\ \bibinfo {author} {\bibfnamefont
  {R.}~\bibnamefont {Singh}},\ }\href@noop {} {\bibfield  {journal} {\bibinfo
  {journal} {Adv. Optical Mater.}\ }\textbf {\bibinfo {volume} {3}},\ \bibinfo
  {pages} {1537} (\bibinfo {year} {2015})}\BibitemShut {NoStop}%
\bibitem [{\citenamefont {Manjappa}\ \emph {et~al.}(2015)\citenamefont
  {Manjappa}, \citenamefont {Chiam}, \citenamefont {Cong}, \citenamefont
  {Bettiol}, \citenamefont {Zhang},\ and\ \citenamefont {Singh}}]{Manjappa15}%
  \BibitemOpen
  \bibfield  {author} {\bibinfo {author} {\bibfnamefont {M.}~\bibnamefont
  {Manjappa}}, \bibinfo {author} {\bibfnamefont {S.-Y.}\ \bibnamefont {Chiam}},
  \bibinfo {author} {\bibfnamefont {L.}~\bibnamefont {Cong}}, \bibinfo {author}
  {\bibfnamefont {A.~A.}\ \bibnamefont {Bettiol}}, \bibinfo {author}
  {\bibfnamefont {W.}~\bibnamefont {Zhang}}, \ and\ \bibinfo {author}
  {\bibfnamefont {R.}~\bibnamefont {Singh}},\ }\href {\doibase
  10.1063/1.4919531} {\bibfield  {journal} {\bibinfo  {journal} {Applied
  Physics Letters}\ }\textbf {\bibinfo {volume} {106}},\ \bibinfo {pages}
  {181101} (\bibinfo {year} {2015})}\BibitemShut {NoStop}%
\bibitem [{\citenamefont {Srivastava}\ \emph {et~al.}(2015)\citenamefont
  {Srivastava}, \citenamefont {Manjappa}, \citenamefont {Cong}, \citenamefont
  {Cao}, \citenamefont {Al-Naib}, \citenamefont {Zhang},\ and\ \citenamefont
  {Singh}}]{Srivastava15}%
  \BibitemOpen
  \bibfield  {author} {\bibinfo {author} {\bibfnamefont {Y.~K.}\ \bibnamefont
  {Srivastava}}, \bibinfo {author} {\bibfnamefont {M.}~\bibnamefont
  {Manjappa}}, \bibinfo {author} {\bibfnamefont {L.}~\bibnamefont {Cong}},
  \bibinfo {author} {\bibfnamefont {W.}~\bibnamefont {Cao}}, \bibinfo {author}
  {\bibfnamefont {I.}~\bibnamefont {Al-Naib}}, \bibinfo {author} {\bibfnamefont
  {W.}~\bibnamefont {Zhang}}, \ and\ \bibinfo {author} {\bibfnamefont
  {R.}~\bibnamefont {Singh}},\ }\href {\doibase 10.1002/adom.201500504}
  {\bibfield  {journal} {\bibinfo  {journal} {Advanced Optical Materials}\
  }\textbf {\bibinfo {volume} {4}},\ \bibinfo {pages} {457} (\bibinfo {year}
  {2015})}\BibitemShut {NoStop}%
\bibitem [{\citenamefont {Fedotov}\ \emph {et~al.}(2007)\citenamefont
  {Fedotov}, \citenamefont {Rose}, \citenamefont {Prosvirnin}, \citenamefont
  {Papasimakis},\ and\ \citenamefont {Zheludev}}]{Fedotov07}%
  \BibitemOpen
  \bibfield  {author} {\bibinfo {author} {\bibfnamefont {V.~A.}\ \bibnamefont
  {Fedotov}}, \bibinfo {author} {\bibfnamefont {M.}~\bibnamefont {Rose}},
  \bibinfo {author} {\bibfnamefont {S.~L.}\ \bibnamefont {Prosvirnin}},
  \bibinfo {author} {\bibfnamefont {N.}~\bibnamefont {Papasimakis}}, \ and\
  \bibinfo {author} {\bibfnamefont {N.~I.}\ \bibnamefont {Zheludev}},\
  }\href@noop {} {\bibfield  {journal} {\bibinfo  {journal} {Phys. Rev. Lett.}\
  }\textbf {\bibinfo {volume} {99}} (\bibinfo {year} {2007})}\BibitemShut
  {NoStop}%
\bibitem [{\citenamefont {Singh}\ \emph {et~al.}(2010)\citenamefont {Singh},
  \citenamefont {Al-Naib}, \citenamefont {Koch},\ and\ \citenamefont
  {Zhang}}]{Singh:10}%
  \BibitemOpen
  \bibfield  {author} {\bibinfo {author} {\bibfnamefont {R.}~\bibnamefont
  {Singh}}, \bibinfo {author} {\bibfnamefont {I.}~\bibnamefont {Al-Naib}},
  \bibinfo {author} {\bibfnamefont {M.}~\bibnamefont {Koch}}, \ and\ \bibinfo
  {author} {\bibfnamefont {W.}~\bibnamefont {Zhang}},\ }\href@noop {}
  {\bibfield  {journal} {\bibinfo  {journal} {Opt. Express}\ }\textbf {\bibinfo
  {volume} {18}},\ \bibinfo {pages} {13044} (\bibinfo {year}
  {2010})}\BibitemShut {NoStop}%
\bibitem [{\citenamefont {Singh}\ \emph
  {et~al.}(2011{\natexlab{b}})\citenamefont {Singh}, \citenamefont {Al-Naib},
  \citenamefont {Yang}, \citenamefont {Roy~Chowdhury}, \citenamefont {Cao},
  \citenamefont {Rockstuhl}, \citenamefont {Ozaki}, \citenamefont
  {Morandotti},\ and\ \citenamefont {Zhang}}]{Singh11a}%
  \BibitemOpen
  \bibfield  {author} {\bibinfo {author} {\bibfnamefont {R.}~\bibnamefont
  {Singh}}, \bibinfo {author} {\bibfnamefont {I.~A.~I.}\ \bibnamefont
  {Al-Naib}}, \bibinfo {author} {\bibfnamefont {Y.}~\bibnamefont {Yang}},
  \bibinfo {author} {\bibfnamefont {D.}~\bibnamefont {Roy~Chowdhury}}, \bibinfo
  {author} {\bibfnamefont {W.}~\bibnamefont {Cao}}, \bibinfo {author}
  {\bibfnamefont {C.}~\bibnamefont {Rockstuhl}}, \bibinfo {author}
  {\bibfnamefont {T.}~\bibnamefont {Ozaki}}, \bibinfo {author} {\bibfnamefont
  {R.}~\bibnamefont {Morandotti}}, \ and\ \bibinfo {author} {\bibfnamefont
  {W.}~\bibnamefont {Zhang}},\ }\href {\doibase 10.1063/1.3659494} {\bibfield
  {journal} {\bibinfo  {journal} {Applied Physics Letters}\ }\textbf {\bibinfo
  {volume} {99}},\ \bibinfo {pages} {201107} (\bibinfo {year}
  {2011}{\natexlab{b}})}\BibitemShut {NoStop}%
\bibitem [{\citenamefont {Koshelev}\ \emph
  {et~al.}(2018{\natexlab{b}})\citenamefont {Koshelev}, \citenamefont
  {Lepeshov}, \citenamefont {Liu}, \citenamefont {Bogdanov},\ and\
  \citenamefont {Kivshar}}]{Koshelev}%
  \BibitemOpen
  \bibfield  {author} {\bibinfo {author} {\bibfnamefont {K.}~\bibnamefont
  {Koshelev}}, \bibinfo {author} {\bibfnamefont {S.}~\bibnamefont {Lepeshov}},
  \bibinfo {author} {\bibfnamefont {M.}~\bibnamefont {Liu}}, \bibinfo {author}
  {\bibfnamefont {A.}~\bibnamefont {Bogdanov}}, \ and\ \bibinfo {author}
  {\bibfnamefont {Y.}~\bibnamefont {Kivshar}},\ }\href {\doibase
  10.1103/PhysRevLett.121.193903} {\bibfield  {journal} {\bibinfo  {journal}
  {Phys. Rev. Lett.}\ }\textbf {\bibinfo {volume} {121}},\ \bibinfo {pages}
  {193903} (\bibinfo {year} {2018}{\natexlab{b}})}\BibitemShut {NoStop}%
\bibitem [{\citenamefont {Al-Naib}\ \emph {et~al.}(2012)\citenamefont
  {Al-Naib}, \citenamefont {Singh}, \citenamefont {Rockstuhl}, \citenamefont
  {Lederer}, \citenamefont {Delprat}, \citenamefont {Rocheleau}, \citenamefont
  {Chaker}, \citenamefont {Ozaki},\ and\ \citenamefont
  {Morandotti}}]{AlNaib12}%
  \BibitemOpen
  \bibfield  {author} {\bibinfo {author} {\bibfnamefont {I.}~\bibnamefont
  {Al-Naib}}, \bibinfo {author} {\bibfnamefont {R.}~\bibnamefont {Singh}},
  \bibinfo {author} {\bibfnamefont {C.}~\bibnamefont {Rockstuhl}}, \bibinfo
  {author} {\bibfnamefont {F.}~\bibnamefont {Lederer}}, \bibinfo {author}
  {\bibfnamefont {S.}~\bibnamefont {Delprat}}, \bibinfo {author} {\bibfnamefont
  {D.}~\bibnamefont {Rocheleau}}, \bibinfo {author} {\bibfnamefont
  {M.}~\bibnamefont {Chaker}}, \bibinfo {author} {\bibfnamefont
  {T.}~\bibnamefont {Ozaki}}, \ and\ \bibinfo {author} {\bibfnamefont
  {R.}~\bibnamefont {Morandotti}},\ }\href {\doibase 10.1063/1.4745790}
  {\bibfield  {journal} {\bibinfo  {journal} {Applied Physics Letters}\
  }\textbf {\bibinfo {volume} {101}},\ \bibinfo {pages} {071108} (\bibinfo
  {year} {2012})}\BibitemShut {NoStop}%
\bibitem [{\citenamefont {Gupta}\ \emph {et~al.}(2016)\citenamefont {Gupta},
  \citenamefont {Savinov}, \citenamefont {Xu}, \citenamefont {Cong},
  \citenamefont {Dayal}, \citenamefont {Wang}, \citenamefont {Zhang},
  \citenamefont {Zheludev},\ and\ \citenamefont {Singh}}]{Gupta16}%
  \BibitemOpen
  \bibfield  {author} {\bibinfo {author} {\bibfnamefont {M.}~\bibnamefont
  {Gupta}}, \bibinfo {author} {\bibfnamefont {V.}~\bibnamefont {Savinov}},
  \bibinfo {author} {\bibfnamefont {N.}~\bibnamefont {Xu}}, \bibinfo {author}
  {\bibfnamefont {L.}~\bibnamefont {Cong}}, \bibinfo {author} {\bibfnamefont
  {G.}~\bibnamefont {Dayal}}, \bibinfo {author} {\bibfnamefont
  {S.}~\bibnamefont {Wang}}, \bibinfo {author} {\bibfnamefont {W.}~\bibnamefont
  {Zhang}}, \bibinfo {author} {\bibfnamefont {N.~I.}\ \bibnamefont {Zheludev}},
  \ and\ \bibinfo {author} {\bibfnamefont {R.}~\bibnamefont {Singh}},\ }\href
  {\doibase 10.1002/adma.201601611} {\bibfield  {journal} {\bibinfo  {journal}
  {Advanced Materials}\ }\textbf {\bibinfo {volume} {28}},\ \bibinfo {pages}
  {8206} (\bibinfo {year} {2016})}\BibitemShut {NoStop}%
\bibitem [{\citenamefont {Al-Naib}\ \emph {et~al.}(2015)\citenamefont
  {Al-Naib}, \citenamefont {Yang}, \citenamefont {Dignam}, \citenamefont
  {Zhang},\ and\ \citenamefont {Singh}}]{Al-Naib:15}%
  \BibitemOpen
  \bibfield  {author} {\bibinfo {author} {\bibfnamefont {I.}~\bibnamefont
  {Al-Naib}}, \bibinfo {author} {\bibfnamefont {Y.}~\bibnamefont {Yang}},
  \bibinfo {author} {\bibfnamefont {M.~M.}\ \bibnamefont {Dignam}}, \bibinfo
  {author} {\bibfnamefont {W.}~\bibnamefont {Zhang}}, \ and\ \bibinfo {author}
  {\bibfnamefont {R.}~\bibnamefont {Singh}},\ }\href@noop {} {\bibfield
  {journal} {\bibinfo  {journal} {Appl. Phys. Lett.}\ }\textbf {\bibinfo
  {volume} {106}} (\bibinfo {year} {2015})}\BibitemShut {NoStop}%
\bibitem [{\citenamefont {Doeleman}\ \emph {et~al.}(2018)\citenamefont
  {Doeleman}, \citenamefont {Monticone}, \citenamefont {den Hollander},
  \citenamefont {Al{\`u}},\ and\ \citenamefont {Koenderink}}]{Doeleman2018}%
  \BibitemOpen
  \bibfield  {author} {\bibinfo {author} {\bibfnamefont {H.~M.}\ \bibnamefont
  {Doeleman}}, \bibinfo {author} {\bibfnamefont {F.}~\bibnamefont {Monticone}},
  \bibinfo {author} {\bibfnamefont {W.}~\bibnamefont {den Hollander}}, \bibinfo
  {author} {\bibfnamefont {A.}~\bibnamefont {Al{\`u}}}, \ and\ \bibinfo
  {author} {\bibfnamefont {A.~F.}\ \bibnamefont {Koenderink}},\ }\href
  {\doibase 10.1038/s41566-018-0177-5} {\bibfield  {journal} {\bibinfo
  {journal} {Nature Photonics}\ }\textbf {\bibinfo {volume} {12}},\ \bibinfo
  {pages} {397} (\bibinfo {year} {2018})}\BibitemShut {NoStop}%
\end{thebibliography}%

\end{document}